\documentstyle[12pt]{article}
\newcommand{\drm}{{\rm d}}
\newcommand{\h}{\hspace*{5 ex}}
\newcommand{\vs}{\vspace*}
\newcommand{\disp}{\displaystyle}
\newcommand{\bi}{\bibitem}

\begin{document}

\centerline{\bf ON THE PHENOMENOLOGY OF TACHYON RADIATION. $^{(*)}$}

\footnotetext{$^{(*)}$ Work partially supported by
INFN--Sezione di Catania, MURST and CNR (Italy), and by CNPq (Brazil).}

\vspace*{0.9 cm}

\centerline{Ron Folman}

\vspace*{0.3 cm}

{\small
\centerline{{\em Physics Department, Weizmann Institute of Science, Rehovot,
Israel.}}
}

\vspace*{0.3 cm}

\centerline{ and }

\vspace*{0.3 cm}

\centerline{Erasmo Recami}

\vspace*{0.3 cm}

{\small
\centerline{{\em Facolt\`{a} di Ingegneria, Universit\`{a} Statale di
Bergamo, Dalmine (BG), Italy;}}
\centerline{{\em I.N.F.N., Sezione di Milano, Milan, Italy; \ and}}
\centerline{{\em Dept.$\!$ of Applied Mathematics, State University at
Campinas, Campinas, S.P., Brazil.}}
}

\vspace*{1.5 cm}

{\bf Abstract  \  --} \ We present a brief overview of the different kinds
of electromagnetic radiations expected to come from (or to be induced by)
space-like sources (tachyons). \ New domains of radiation are here considered;
and the possibility of experimental observation of tachyons via electromagnetic
radiation is discussed.

\vspace*{2. cm}



\newpage

{\bf 1. \ Introduction \ --} \ The theoretical possibility of existence of
Superluminal ($V^2 > c^2$) particles has been discussed thoroughly and seems
to need no introduction.$^1$  Recently, this topic received new impulse by
a series of interesting experimental results, about the seemingly
Superluminal speeds associated {\em both} with evanescent waves (let us
recall the Cologne, Berkeley and Florence results),$^2-4$ {\em and} with
electron neutrinos$^5$ and muon neutrinos.$^6$

\h In the past, all experiments performed with
the aim of detecting tachyons had reported ---in practice--- negative
results;$^7$ but those searches were mainly devoted to look for the
electro-magnetic (EM) emission signature of tachyons, on the basis of poor
theoretical assumptions.\ In the following, we first define the different
forms of expected EM radiation due to tachyons, and then re-examine previous
conclusions about tachyon radiations. Finally, we discuss a possible
framework for future experiments. \ We shall always confine ourselves within
the realm of (relativistic) classical physics. \ Below, we shall call
``bradyon" any sub-luminal ($v^2 < c^2$) particle.

\vs{1. cm}

{\bf 2. \ Radiation definitions \ --} \ Since it seems that clear definitions
are still lacking with regard to the possible forms of EM radiation, let us
define what follows:\hfill\break

{\em Definition a):\/} \ $<<$ Let us call IER any EM radiation {\em induced
from a
medium} by a particle (bradyon or tachyon), i.e., coming from the medium after
that the particle has excited it.$>>$ \ This radiation can a priori be conic
(``coherent" ) or not. \ Defined as such, this radiation cannot be induced by
any particle (in particular, by any tachyon) in vacuum  [unless one assumes
a suitable vacuum structure, passing to quantum field theories].\hfill\break

{\em Definition b):\/} \ $<<$ Let us call CER ({\em Cherenkov} EM radiation)
any
IER provoked by a particle which is outside its ``allowed speed--region":
i.e., by a bradyon endowed with a speed $w$ larger than the light speed $v$
in the
medium, or by a tachyon which happens to possess a speed $W$ smaller than the
light speed $V$ in a
suitable medium.$>>$ \ Notice that also CER is not necessarily conic, a priori.
\hfill\break

{\em Definition c):\/} \ $<<$ Let us call OER ({\em ordinary} EM radiation) any
radiation emitted directly by the particle itself, due to its acceleration
or deceleration.$>>$ \ Defined as such, this radiation can be emitted by a
particle (e.g., by a tachyon) also in vacuum, of course. \ In this note, we
shall not deal with this kind of radiation [for details cf. refs.$^1$]. \
Here, let us only recall the following, particular cases:\hfill\break

{\em case A1\/}) CER(B): in a medium with proper refraction index $n_0$, we may
have CER
from a bradyon B when $v < w < c$, that is to say, only when $n_0 > 1$; in
this case the CER is {\em conic\/};\hfill\break

{\em case A2\/}) CER(T): in a medium with proper refraction index $n_0$, we may
have CER
from a tachyon T when $c < W < V$, that is to say, only when $n_0 < 1$; in
this case the CER will be {\em non-conic\/}, since the tachyon position will
always remain inside the (spherical) waves;\hfill\break

while:\hfill\break

{\em case B1\/}) OER(B): in vacuum (for instance), the OER from a bradyon B is
{\em non-conic\/};\hfill\break

{\em case B2\/}) OER(T): in vacuum (for instance), the OER from a tachyon T
will be on the contrary {\em conic\/}.

\vs{1. cm}

{\bf 3. \ Tachyon radiation domains \ --} \ Recami and Mignani$^1$ observed
that the experimental searches for (tachyonic) CER(T)
had probably failed due to the mere fact that no such
radiation is expected to be emitted by the vacuum, nor by {\em normal}
sub-luminal media. \ They overlooked, however, that ---as we saw  for example
under point {\em A2\/}) above--- a CER(T) {\em can} come also from {\em
suitable}
sub-luminal media. \ We want here to discuss such a possibility. \ Let us
specify that below we shall refer ourselves to sub-luminal observers $O$ only,
while a medium cam be either sub-luminal [$u^2 < c^2$] or Super-luminal
[$U^2 > c^2$] with respect to the observer $O$.

\h Recami and Mignani had based themselves on their generalization of
the ``drag effect" for Superluminal speeds, i.e., on their extension of the
{\em apparent} ( = relative to the observer $O$) velocity $v \equiv c/n$ of
light, in a moving medium, for tachyonic media.$^1$ \  The equation for the
``apparent" refraction index $n(u)$ was [both for subluminal and for
Superluminal speeds $u$]:

\

\hfill{
$ n \; \equiv \; n(u) \; = \; {\disp {{n_0 c + u} \over {c + n_0 u}}} \ . $
\hfill} (1)

\

In ref.$^1$, however, only the particular case was considered in which
the proper refraction index $n_0$ of the considered medium is larger than 1
(see our Fig.1, as well as Fig.22 in the first one of
refs.$^1$). \ This led them to claim: \ (i) that a bradyon B is not
expected to ``emit" CER(B) in any Super-luminal media, due to the fact that the
apparent refraction index $n$ in this case is
smaller than 1, thus giving rise to an apparent light speed $v$ in the medium
larger than $c$: a speed that B will never exceed [actually, in Fig.1 we
can see that Super-luminal media possess a refraction index $n$ which is
smaller than 1 and aims asymptotically at $1/n$]; \ and,
because of the Duality Principle:$^1$ \ (ii) that analogously a tachyon T
will not be able to ``emit" CER(T) in any sub-luminal media. \

\h But, of course, media also exist possessing a proper refraction index
$n_0$ {\em smaller} than 1; \ in this case, eq.(1) yields the behaviour
depicted in our Fig.2. \ And from Fig.2 one can infere that in fact
a tachyon T can ``emit" CER(T) in a sub-luminal medium with $n < 1$, provided
that its speed $w$ obeys the constraint \ $c < w < v$, \ where $v = c/n > 1$.
 \ Analogously, because of the Duality Principle, a bradyon B will be able to
``emit" CER(B) in a Super-luminal medium such that $n_0 < 1$ (as can be
derived from Fig.2).

\vs{1. cm}

{\bf 4. \ Experimental considerations. \ --} \ The fact that tachyons T are
now expected to emit CER(T) also in {\em suitable} sub-luminal media (for
speeds ranging between $c$ and $c/n$) has of course some experimental
consequences. \ In fact, we do not know of any Super-luminal media;
 \ on the contrary, sub-luminal media with $n_0
< 1$ are available, e.g., in the form of {\em plasmas}. \ We shall therefore
consider the possible signature of CER(T) in a plasma.

\h Let us first {\em assume} the following: \ Any particle in a medium
will find its way
to loose quickly energy (e.g., by inducing emissions from the medium), when
it happens to be outside its ``allowed speed--region"; namely, in the case
of a tachyon, when its speed $w$ is between $c$ and the apparent light
speed $v$. \ This Assumption seems to be supported also by the observation
that ---in the case of bradyons--- the function $-\drm E / \drm x$  \ {vs.} \
speed presents for most particles an enhancement when crossing the $v = c/n$
line (the so-called ``relativistic rise"). Such an enhancement, incidentally,
is observed in several energy--deposit channels, and not solely in the EM
channel
(which constitutes just a portion of the energy loss).

\h A tachyonic CER(T) in a plasma is expected to be non-coherent, i.e.,
{\em non-conic\/}; and therefore weak. In fact, the interested tachyon has to
move slower than the apparent light speed, so that the standard Cherenkov
geometry is not realysed. \ This does constitute a real difficulty for any
experimental search dedicated to CER(T) in plasmas, and it might seem
preferable to look for {\em conic} IER(T) in the speed range above $c/n$,
if it weren't for our Assumption above. \ Actually, tachyon interactions
with electromagnetic fields and with ordinary matter are not yet completely
known [different effects, such as that a point charge when Super-luminal
is expected to be spread over a double cone,$^{1,8}$ have not yet been
fully dealt with]; and in particular we do not know
how and how much IER would be emitted by a tachyon. In any case, those
tachyon interactions can be predicted to be weak, on the basis of the fact
that a Super-luminal electric charge is predicted to behave as a magnetic
pole$^{1,9}$ [in the sense, roughly speaking, that the intensities of the
electric and magnetic fields generated by a charge get interchanged
between themselves when passing from the speed $w$ to the ``dual" speed
$W = c^2/w$].$^{10}$ \ Therefore, in force of our Assumption, we are
entitled to consider as theoretically reasonable a search for CER(T), in
the speed range $c$ to $c/n$.

\h We might then suggest the use of an
array of detectors, situated parallel to the tachyonic flight path,
in order to try to observe the advance of the Super-luminal source. \ Another
suggestion would be to confine the plasma within a $n > 1$ substance (the
border between the two substances being parallel to the expected tachyonic
track and close to it), in which a part of the  radiation circles emitted
by the plasma will form a coherent Cherenkov front with angles larger than
those permitted for bradyon sources.

\vs{1. cm}

{\bf 5. \ Conclusions and discussion \ --} \ After having mentioned some
recent experimental results, that might indicate the existence of
Superluminal objects, we revised and corrected in this note the theoretical
reasons why  it seemed till now that
direct detection of Cherenkov radiation, induced by tachyons in ordinary
(subluminal) media, was impossible.

\h Then, we presented arguments favouring ---on the contrary---  the
{\em possibility} of searching
for  Cherenkov emission induced by tachyons {\em in} (subluminal)
{\em plasmas}.

\h We wish to conclude with the following discussion.

\h First of all,
one should not forget that,
as mentioned above, while the electric
interaction constitutes the major interaction betwen a charged bradyon and
matter, on the contrary the magnetic interaction will possibly be the
dominant one for a charged tachyon interacting with ordinary
matter.$^{1,9,10}$ \ So
that, generally speaking, one ought to try to detect Super-luminal sources
via the magnetic
field they are supposed to generate. \ More in general, it would be quite
useful to solve ---before all--- the Maxwell equations suitably generalized for
Superluminal charges, in order to find out the electric and magnetic field
that a charged tachyon is expected to create in space-time; since such a
study (preliminary performed in ref.$^{11}$, at an elementary level) will
indicate how a Superluminal charge is expected to interact with
ordinary matter. \ Let us recall that two different generalizations for Maxwell
equations have been proposed: cf. refs.$^{1,10}$ and refs.$^{12}$,
respectively.

\h Second, the question of wherefrom
(i.e., from what sources) tachyons
are expected to come is still an open problem; even if different sources
(including cosmic showers$^{13}$) have been suggested.$^{1,14}$

\h At last, let
us remark
that another interesting search could be the one devoted to detect the
(conic) OER(T) emitted by tachyons ---in vacuum, for instance--- when
accelerating because of {\em bremsstrahlung}--type processes.

\vs{1. cm}

\vspace*{1.7 cm}

{\bf Acknowledgements:} This brief note is dedicated to the memory of the
late Professor W. Yourgrau and the many discussions we had with him (and
T.K. Shah)
on similar subjects at Trieste about fifteen years ago. \ One of us (R.F.)
would like to thank M.T. Teli
for enlightening discussions; while the second author (E.R.) acknowledges the
kind collaboration of A. Bugini, R. Garattini, L. Lo Monaco, R. Monaco,
E.C. de Oliveiras,
M. Pignanelli, G.M. Prosperi, W.A. Rodrigues, S. Sambataro, P. Saurgnani,
M.T. Vasconselos
and particularly F. Raciti and G. Salesi.

\newpage

\centerline{{\bf Figure Captions}}

\vspace*{1. cm}

{\bf Fig.1} --  The ``drag effect" behaviour considered by Recami and
Mignani,$^1$ on the basis of equation (1).  \ It refers to the case in which
the proper refraction index of the considered medium is $n_0 > 1$.

\vspace*{1. cm}

{\bf Fig.2} -- The ``drag effect" complementary  behaviour [still derived
from eq.(1)]  for the case
in which the proper refraction index of the considered medium is $n_0 < 1$.
 \ This case had been overlooked in ref.$^1$. \ For the possible
 consequences, see the text.

\end{document}